**P-Values in a Post-Truth World**

Joshua T. Vogelstein, Johns Hopkins University

**The role of statisticians in society is to provide tools, techniques, and guidance with regards to how much to trust data. This role is increasingly more important with more data and more misinformation than ever before. The American Statistical Association recently released two statements on p-values, and provided four guiding principles. We evaluate their claims using these principles and find that they failed to adhere to them. In this age of distrust, we have an opportunity to be role models of trustworthiness, and responsibility to take it.**

It is widely believed that we currently live in a post-truth world. Evidence cited includes the emergence of "fake news" [1], the frequency of false or misleading statements from world leaders [2], social media companies' refusal to regulate misleading political ads [3], the reproducibility crisis [4], and rampant misinformation about Coronavirus Disease 2019 (COVID-19) [5]. Theories of the causal factors leading to this situation include continual false promises from politicians, pundits, and other experts, the public having had "enough with experts" [6], and scientific publications with misleading or inaccurate results [7].

This state of affairs should be, and is, deeply troubling to statisticians [8]. Statistics can reasonably be thought of as a kind of quantitative epistemology. The epistemological belief underlying statistics mirrors philosophical ideas elaborated in "Two Dogmas of Empiricism", in which Quine points out that there is no real distinction between analytic and synthetic truths, or, in other words, there is no Truth (with a capital T) [9]. Statistics, rather than searching for Truth, builds models of the world. George Box, a leading statistician of the 20th century, famously quipped, "All models are wrong, some are useful [10]." In this sense, statistics enables the evolution of ideas [11], much like selective pressures enable the evolution of species. Although "survival of the fittest" is a typical characterization of selective pressure in evolution, a more accurate description is "survival of the fit enough". In other words, evolution is not trending towards the most fit species ever. Rather, species continue to evolve as niches change. Similarly, in statistics, it is not the "survival of the fittest" model, but rather, "survival of the fit enough." Every experiment and every analysis is (at least implicitly) a model comparison, and the hope is to either gather evidence in favor of the existing model, or identify discrepancies from the model to improve, refine, or replace it. *The role of statistics in society is to provide selective pressure to existing models, weeding out those that are not deemed fit enough.*

It is the responsibility of the statistics community, therefore, to be epistemological role models. The key to validly assessing the fitness of a model rests in both the acquisition and analysis of data. Statisticians have formalized these processes, combining prior knowledge with data to build evidence for or against any given hypothesis. In fact, we have developed many tools, with competing assumptions, and sometimes different conclusions [12]. All approaches, however, begin with designing an experiment [13], and proceed to acquire data, followed by analyzing that data [14].

It is in this context that the American Statistical Association (ASA) has decided to end its silence with regard to its perspective on statistical inference. In particular, ASA released a statement in 2016 discussing the perils of p-values [15] (which we refer to hereafter as ASA I, following Deborah Mayo's convention [16]. More recently, the ASA took a further step, publishing an editorial recommending the elimination of the dichotomization of p-values, that is, they recommended that the language "statistically significant", essentially, be banned [17] (which we refer to as ASA II). They provide a number of reasons for their conclusions, as well as some value-based guidance on how to proceed. ASA II provides a mnemonic, ATOM, which summarizes its values:

- Accept uncertainty,
- be Thoughtful,
- be Open, and
- be Modest.

The values espoused by the ASA seem appropriate for making any argument, whether it is statistical or not. We therefore evaluate the ASA statements on p-values, in light of these values. Concomitant with the publication of ASA II, the American Statistician also published 43 commentaries, and Nature published a comment called "Retire Statistical Significance" (we refer to this as AGM hereafter) [18], which continues to be the publication with the highest Altmetric score ever to-date [19]. The main question that motivates both ASA II and AGM is whether the use of the term statistical significance accelerates or impedes scientific progress. This question, like any other empirical question, is one that can best be answered by designing a thoughtful experiment, carefully acquiring data, and honestly analyzing it.

The central thesis of this document is that the ASA II does not reflect best practices with regards to either empiricism or its espoused values: no experiments to evaluate its claims are conducted or provided, nor does it follow the ATOM guidelines. It may continue to be socially acceptable for politicians or pundits to argue without quantitative evidence in support of their claims, without accepting uncertainty, being thoughtful, open, or modest. But statisticians can be held to a higher epistemological standard, a standard that we invented and continue to refine.

Below, I delve further into the four values that ASA II espouses. For each of the four values, we state the value, followed by a quote from ASA II (which is itself typically a quote from one of the 43 presented commentaries) that summarizes the value, and then evaluate the recommendations of ASA II in light of that value. I then propose an ASA III, which can be a model of living up to these values.

## 1. Accept uncertainty

So it is time to…"move toward a greater acceptance of uncertainty and embracing of variation" [20].

Section 2 of ASA II is titled, "Don't Say 'Statistically Significant.'" Presumably, the authors are relatively certain that stating "statistical significance" is net negative. The section concludes blithely with, "In sum, 'statistically significant'—don't say it and don't use it." No acceptance of uncertainty or embracing of variation is implied or stated in this section, with regards to whether/when one can/should say the phrase. This is despite the fact that many of the commentaries on both ASA I and ASA II, jointly published, express uncertainty of this conclusion, and embrace variation. In fact, Section 5 of ASA II includes several quotes expressing both greater uncertainty and an embrace of variation. For example, Greenland is quoted as saying [21]:

> The core human and systemic problems are not addressed by shifting blame to p-values and pushing alternatives as magic cures—especially alternatives that have been subject to little or no comparative evaluation in either classrooms or practice, implying the need for empirical investigations of these claims, rather than blanket statements. Section 5 also quotes Fricker et al. [22], which empirically evaluated the impact of the journal *Basic and Applied Social Psychology* banning all of inferential statistics, including the phrase "statistically significant." That study found that when authors were forbidden from using statistical inference, studies tended to overstate, rather than understate their claims, providing evidence that some proposals had the opposite of the desired impact. Section 5 also acknowledges that after asking authors in the special issue to identify when p-value thresholds are permissible, they provided four distinct types of scenarios.

Perhaps the most interesting aspect of the recommendation of ASA II is to report continuous (i.e., exact) p-values, rather than dichotomizing:

> Where p-values are used, they should be reported as continuous quantities (e.g., p = 0.08).

Interestingly, reporting an exact p-value is antithetical to the ethic of accepting uncertainty for a number of reasons. First, p-values can rarely be computed exactly. This is because of the definition of p-values: the probability (under the null) of observing a test statistic more extreme than the observed test statistic. To *exactly* compute this null distribution requires being able to compute it for the particular observed sample size. However, it is extremely rare that we have a closed-form solution to this problem. Whenever we lack closed-form solutions for the null distribution of the test statistic for a given sample size, we must *approximate* the p-value, typically using one of two approaches.[1] The first approach approximates the distribution of the test statistic under the null *asymptotically*. This approach results in essentially all classical statistical tests, including t-tests, ANOVA, etc. This is an approximation because we never have infinite data in practice. The second approach approximates the distribution of the test statistic under the null using resampling approaches, such as bootstrap or permutation. There is no finite number of bootstraps one can sample to get an exact p-value. Similarly, the number of permutations one must compute to obtain an exact p-value is typically astronomically, super-exponentially high (except for a few cases, such as Fisher's exact test). Specifically, it is the number of permutations of the number of samples ("n!"). In practice, analysts compute a tiny fraction of the possible permutations. Perhaps the authors of ASA II did not mean "exact", but rather, "don't dichotomize". There is a spectrum from exact to dichotomized, and if they meant to choose somewhere on the spectrum, further guidance would be desirable.

---

[1] Even if we did not have to approximate computing the p-value, we would still have to *report* approximate p-values in most cases. This is because the p-value, in general, is not a number with only a few significant figures (in fact, it is not necessarily even a rational number).

In light of all this, a short summary of what Section 2 might have said, had it followed the dictum of accepting uncertainty, is this:

> We believe that it is typically the case that stating that a result is "statistically significant" is net negative with regard to scientific progress, meaning that it will either increase the likelihood of reporting a false discovery, or decrease the likelihood of finding a true discovery, all else being equal. Moreover, we believe that all models are wrong, and some are useful; therefore, all null models are false, so statistical significance can simply reflect sample size. The direct empirical evidence in support of our belief of the net negativity of dichotomizing results on the basis of p-value thresholds is limited. Nonetheless, we recommend, instead, reporting an approximate p-value, such as, p ≈ 0.08 (and we recommend reporting only a few significant digits determined by sample size). We acknowledge that there are situations for which stating that a result is statistically significant is helpful, such as industrial quality control and pre-registered medical trials, assuming the inferences are conducted honestly and appropriately. We do not propose guidelines for when we deem stating a result is statistically significant is net positive, but rather, suggest that future empirical investigations are warranted to determine suitable concrete guidelines.

## 2. Be thoughtful

> Statistically thoughtful researchers begin above all else with clearly expressed objectives…. "[In thoughtful research] modeling assumptions (including all the choices from model specification to sample selection and the handling of data issues) should be sufficiently documented so independent parties can critique, and replicate, the work" [23]…. "Thoughtful research prioritizes sound data production by putting energy into the careful planning, design, and execution of the study" [24]. Thoughtful research includes careful consideration of the definition of a meaningful effect size.

It is widely acknowledged that the most difficult component of research is in asking the right questions, not making inferences. What is the right question to ask here? One option would be:

is computing p-values and/or stating that results are statistically significant net positive or net negative for science? Profitably answering this question, like most others, warrants being thoughtful, as described above. What is the appropriate notion of effect size to characterize the scientific value of p-values and statistical significance? What is a reasonable experimental design? Who should conduct this experiment, and how should the results be reported such that interested independent parties can critique, replicate, and extend this work? None of these questions are asked or answered in ASA I or II or AGM.

A related research question would be: which of the many proposed alternatives to reporting p-values and/or statistical significance would be most beneficial to science [25]? Thoughtfully answering this question would benefit from understanding the primary causal factors leading to the inefficacy of reporting statistical significance. For example, if the main source of difficulty is in *understanding* p-values, then any approach that supplements p-values with additional statistical quantities is not likely to resolve the problem, as then individuals may need to understand both p-values and something else.

Based on this, a paragraph to follow the first proposed paragraph, in alignment with the value of thoughtfulness, may read like this:

> A key next step to understand the trustworthiness of various data analytic approaches is to further study them, for example, by studying reproducibility of results across varying inferential techniques, including p-values and statements of significance. We therefore call on various funding mechanisms, including government and private funding sources, to announce mechanisms to study the trustworthiness of different strategies. Such funding mechanisms will provide a much needed incentive to researchers to design studies which investigate these issues. To increase the reproducibility of the resulting studies, we suggest they include a mandatory pre-registration and/or validation dataset. This will guarantee that effect size is meaningfully defined *a priori*, and that interested independent parties can critique and replicate the work.

## 3. Be Open

> "[W]e should base judgments on *evidence* and careful reasoning, and seek wherever possible to eliminate potential sources of bias" [26].....To be open, remember that one study is rarely enough.

Evidence supporting the net negativity of claims of statistical significance is largely missing from ASA I and II, perhaps because the authors believe the evidence is either "self-evident" or otherwise available (though some evidence suggests the opposite [22]). And while careful reasoning is subjective, it has been extensively studied. One particularly salient effort related to this is the *Good Judgement Project*, spearheaded by Professor Phillip Tetlock.

However, evidence in support of the proposed alternatives, or even to support discontinuing the use of "statistically significant," is lacking from ASA I and II (AGM does provide empirical evidence of wrong interpretations of significance in the literature). Moreover, potential sources of bias are not acknowledged, much less eliminated, in either document. "Expert Political Judgment: How Good Is It" is a book that summarizes Tetlock's work over 20 years of asking people to make predictions [27]. He concludes that most categorizations are not useful for distinguishing people into groups of those who make predictions better than chance, and those who do not. This includes categorizations based on level of education or expertise, political persuasion, etc. Tetlock does find, however, that expert judgement does depend on one's ability to contemplate counterfactuals, or potential outcomes. That is, individuals who expressed both sides of an argument were more likely to make accurate predictions than those who only expressed rationale in favor of their side of the argument. This process of contemplating counterfactuals is arguably necessary, though not sufficient, to consider an argument "carefully reasoned," and therefore open to being wrong. Neither ASA II nor AGM provided any counterfactuals (AGM acknowledged the possibility of counterfactuals). Had ASA II acknowledged counterfactuals, it may have included a paragraph such as:

> One study is rarely enough, and this holds for the scientific/social impact of stating results are statistically significant. This is a further reason for funders to create mechanisms to study these issues. While we believe that stating results are statistically significant is net negative, this view is not universal. Indeed, statistical significance claims rose, at least in part, to combat overstated claims of effect. Moreover, the concept of statistical significance is sufficiently broad, theoretically justified, and principled, such that it has been widely adopted and used for nearly a century. In fact, essentially all FDA-approved medical practices were approved on the basis of a statistically significant, pre-registered clinical trial. As Benjamini [28] stated, statistical significance "offers a first line of defense against being fooled by randomness, separating signal from noise," and has thus been incredibly useful in a wide diversity of fields. In that sense, its scientific value has been extremely high, even though it also has had considerable costs. Perhaps these costs are due not to the statement of statistical significance, but rather, its misinterpretation and misuse.

## 4. Be Modest

> Be modest about the role of statistical inference in scientific inference…. "Scientific inference is a far broader concept than statistical inference" [29]... Because of the strong desire to inform and be informed, there is a relentless demand to state results with certainty…. Resist the urge to overreach in the generalizability of claims…. Accept that both scientific inference and statistical inference are hard, and understand that no knowledge will be efficiently advanced using simplistic, mechanical rules and procedures.

Donald Rubin, a famous statistician, quipped, "No causation without manipulation." And yet, it seems that part of the premise of ASA I and ASA II is that a substantial contributing factor to the distrust of analytical results is the use of post-acquisition data analysis techniques and claims. Being modest about the role of statistical inference in scientific inference goes both ways. Another result from Tetlock's Good Judgement Project is the identification of a second quality that people who make accurate judgements have: modesty. More specifically, they update their posteriors given new data, rather than assuming the data are wrong in some fashion. In light of this, ASA III could include the following paragraph:

> The causal factors underlying the distrust are undoubtedly myriad and multifarious. We believe that data analysis procedures are at least partially to blame. That said, poor data measurement, low sample sizes, lack of validation datasets, and existing incentive structures likely play a contributory role. Experiments in which we vary only the data analysis could resolve these issues, and so we advocate for such experiments to be funded and conducted. If the data indicate that whether or not one writes "statistically significant" is not meaningful, or does not play a large negative role in the replication crisis, our recommendation was wrong. On the other hand, experiments may find the main difference between (i) results that replicate and (ii) those that do not, is that those that replicate always have a validation study. If that were the case, or something like it, it could focus our energy on developing new and more efficient ways of obtaining validation data [30], which would be a very different focus than discussing the relative merits and demerits of p-values and significance. Regardless, in light of being modest, we recommend that every scientific study reports approximately one hypothesis, and therefore, reports approximately one p-value, quantified using a registered analysis procedure on a validation dataset. For those without such a validation dataset, we recommend reporting results as exploratory (which they are) rather than confirmatory (which they are not).

## 4. Concluding Thoughts

To repeat Greenland's [21] summary,

> The core human and systemic problems are not addressed by shifting blame to p-values and pushing alternatives as magic cures—especially alternatives that have been subject to little or no comparative evaluation in either classrooms or practice... What we need now is to move beyond debating only our methods and their interpretations, to concrete proposals for elimination of systemic problems.

We agree. Perhaps more importantly, we, as statisticians, can take this opportunity to be role models, reach conclusions ideally after careful evidence-based analysis, and make policy recommendations also only after careful policy analysis. Trust in scientists is on the rise (Trust and Mistrust in Americans' View...). The ongoing COVID-19 pandemic makes our responsibility of being trustworthy all the more urgent and important, because only by virtue of the application of statistical tools will we as a global society be able to determine when we are safe again.

## Acknowledgements

We would like to thank Carey E. Priebe, the NeuroData team, and Andrew Gelman for helpful comments. And thank you for trusting me with your time by reading this article.